\documentstyle[12pt,supercite]{article}

\def\sqr#1#2{{\vcenter{\hrule height.#2pt
      \hbox{\vrule width.#2pt height#1pt \kern#1pt
        \vrule width.#2pt}
      \hrule height.#2pt}}}

\def\abstracts#1#2#3{{
        \centering{\begin{minipage}{4.62in}\baselineskip=13pt
        \small
        \centerline{\bf Abstract}
        \vspace*{0.2cm}                %
        \parindent=0pt #1\par
        \parindent=18pt #2\par
        \parindent=15pt #3
        \end{minipage} }\par}}
\renewcommand{\thefootnote}{\fnsymbol{footnote}}
\begin{document}
\vspace*{-3cm}
\centerline{\LARGE \bf The Ising Model on 3D Random Lattices:}
\vspace{0.2cm}
\centerline{\LARGE \bf A Monte Carlo Study}
\vspace{0.6cm}
\renewcommand{\thefootnote}{\arabic{footnote}}
\centerline{\large {\em Wolfhard Janke\/}$^{1}$
             and   {\em Ramon Villanova\/}$^{2}$}
\vspace{0.4cm}
\centerline{\large $^1$ {\small Institut f\"ur
                                Theoretische Physik,
                                Universit\"at Leipzig,}}
\centerline{{\small Augustusplatz 10/11, 04109 Leipzig, Germany}}
\vspace{0.15cm}
\centerline{\large $^2$ {\small Matem\`{a}tica Aplicada, DEE,
                      Universitat Pompeu Fabra,}}
\centerline{{\small c/ Ramon Trias Fargas, 25-27, 08005 Barcelona,
Spain}}
\vspace{2.50cm}
\abstracts{}{
We report single-cluster Monte Carlo simulations of the Ising model on
three-dimensional Poissonian random lattices with up to $128\,000 \approx 50^3$
sites which are linked together according to the Voronoi/Delaunay prescription.
For each lattice size quenched averages are performed over 96 realizations.
By using reweighting techniques and finite-size scaling analyses we investigate
the critical properties of the model in the close vicinity of the phase
transition point. Our random lattice data provide strong evidence that, for
the available system sizes, the resulting effective critical exponents
are indistinguishable from recent high-precision estimates obtained in
Monte Carlo studies of the Ising model and $\phi^4$ field theory on
three-dimensional regular cubic lattices.
}{}
\thispagestyle{empty}
\newpage
%
     \section{Introduction} \label{sec:intro}
%
Experimental studies of the critical behavior of real materials are
often confronted with the influence of impurities and inhomogeneities.
For a proper interpretation of the measurements it is, therefore,
important to develop a firm theoretical understanding of the effect of
such random perturbations. In many situations the typical time scale of
thermal fluctuations in the idealized ``pure'' system is clearly separated
from the time scale of the impurity dynamics, such that to a very good
approximation the impurities can be treated as quenched (frozen), random
disorder.

The importance of the effect of quenched, random disorder on the critical
behavior of a physical system can be quite generally classified by the
critical exponent of the specific heat of the pure system, $\alpha_p$.
The Harris criterion \cite{harris}
asserts that for $\alpha_p >0$ quenched, random disorder
is a relevant perturbation, leading to a different critical behavior than
in the pure case. In particular one expects \cite{chayes} in the disordered
system that $\nu \ge 2/D$, where $\nu$ is the correlation length exponent and
$D$ the dimension of the system. Assuming hyperscaling to be
valid, this implies $\alpha = 2 - D\nu \le 0$. For $\alpha_p < 0$
disorder is irrelevant, and in the marginal case $\alpha_p = 0$ no
prediction can be made. For the case of (non-critical) first-order phase
transitions it is known that the influence of quenched, random disorder can
lead to a softening of the transition \cite{berker}.

Many theoretical and numerical studies have been devoted to
quenched, random site-dilution (SD), bond-dilution (BD), or more general
random-bond (RB) systems. Since for the three-dimensional (3D) Ising model
it is well known that $\alpha_p \approx 0.1 > 0$, quenched, random
disorder should be relevant for this model. This has indeed been verified
by a variety of techniques: Monte Carlo (MC) simulations for
SD \cite{RB_3Dis,WiDo98,Ba_etal98} and BD \cite{BCBJ_01,BCBJ_02},
high-temperature series (HTS) expansions for BD \cite{HJ_01} and field
theoretical renormalization group studies
\cite{Varnashev2000,holo00,soko00,Pelissetto2000}.
For an excellent review and an extensive list of experimental, theoretical
and numerical estimates in the last two decades, see Ref.~\cite{Folk2001}.
As a result consensus has been reached that, while the critical exponent ratios
$\gamma/\nu = 2 - \eta$ and $\beta/\nu$ are hardly distinguishable from the
pure model, the correlation length exponent $\nu$ clearly signals the
disordered fixed point,
\begin{eqnarray}
&&\!\!\!\!\gamma/\nu = 1.966(6),\ \beta/\nu=0.517(3),\
\nu=0.6304(13) \quad
\mbox{(pure \cite{Guida1998})},\nonumber\\
&&\!\!\!\!\gamma/\nu = 1.970(3),\ \beta/\nu=0.515(2),\
\nu=0.678(10) \quad \,\,\,\mbox{(disordered
\cite{Pelissetto2000})},\nonumber
\end{eqnarray}
and, moreover, satisfies the bound $\nu \ge 2/D = 2/3$ in the disordered
case.

Recently also the predicted softening effect at first-order phase transitions
has been confirmed for 3D $q$-state Potts models with
$q \ge 3$ using MC \cite{Ballesteros00,CBJB_01,CBBJ_02} and
HTS \cite{HJ_lat01} techniques.
The overall picture is even better in two dimensions (2D) where several
models with $\alpha_p > 0$
(SD Baxter model \cite{RB_baxter},
SD Baxter-Wu model \cite{RB_baxter_wu},
3- and 4-state RB Potts model \cite{RB_potts},
Ashkin-Teller model \cite{wiseman}) and the marginal ($\alpha_p = 0$)
2D Ising model
\cite{RB_2Dis,RB_2Dis_th1,RB_2Dis_th2,RB_2Dis_MC,RB_2Dis_HTS}
have been
investigated. Also the softening of the first-order transitions for the 2D
models with $q \ge 5$ has been confirmed \cite{cfl92,celik97,cb98,roder}.
For a recent review, see Ref.~\cite{BeCh02}.

In this paper we study another type of quenched, random disorder, namely
{\em connectivity disorder\/}, a generic property of random lattices
whose local coordination numbers vary randomly from site to site.
Physically the concept of random lattices plays an important role in an
idealized description of the statistical geometry of random packings of
particles \cite{mei,collins,ziman}. A prominent example is the
crystallization process in liquids, and many statistical properties of random
lattices have been studied in this context \cite{itzyk}. From a more
technical point of view, random lattices provide a convenient tool to
discretize spaces of non-trivial topology without introducing defects or
any kind of anisotropy \cite{cfl,jh97,jh99}. This desirable
property
has been exploited in a great variety of fields,
ranging from diffusion limited
aggregation \cite{dla} and growth models for
sandpiles \cite{puhl} over the statistical mechanics of membranes and
strings \cite{membranes} to quantum field theory and quantum
gravity \cite{cfl,jh97,jh99,qft,ren}.
The preserved rotational, or more generally Poincar\'e,
invariance suggests that spin systems or field theories defined on random
lattices should reach the infinite-volume or continuum limit faster than on
regular lattices. Conceptually, however, such an approach does only make
sense
as long as the critical properties of the considered system are
not modified by the irregular lattice structure. In view of the
quite general Harris criterion this is a non-trivial question (in
particular due to the inherent spatial correlations of the
disorder in this case), at least for systems characterized by
$\alpha_p \ge 0$.

Specifically we considered 3D Poissonian random lattices of
Voronoi/De\-launay type, and performed an extensive computer
simulation study of the Ising model for lattices varying in size
from $N = 2\,000 \approx 13^3$ to $128\,000 \approx 50^3$ sites.
For each system size quenched averages over the connectivity
disorder are approximated by averaging over 96 independent
realizations. We concentrated on the close vicinity of the
transition point and applied finite-size scaling (FSS) techniques
\cite{fss} to extract the critical exponents and the (weakly
universal) ``renormalized charges'' $U^*_2$ and $U^*_4$. To
achieve the desired accuracy of the data in reasonable computer
time we applied the single-cluster algorithm \cite{1c,sw} to
update the spins and furthermore made extensively use of the
reweighting technique \cite{fs1}.

Previous studies of connectivity disorder focused mainly on 2D where
pronounced effects were observed in MC simulations of $q$-state Potts models
on quenched, random graphs provided by models of quantum gravity
(modified universality classes for $q=2$
and 4 \cite{bhj94,abt99,jj98}, and softening for $q=10$ \cite{jj98,bjj96}).
For 2D random lattices of Voronoi/Delaunay type, on the other hand, no influence
was seen in simulations for $q=2$ \cite{espriu,jkv93} and $q=8$ \cite{jv95}.
The main difference between these two types of random lattices
is the highly fractal structure \cite{fractal} of random gravity graphs
which suggests a stronger ``degree of disorder''.
A similar dependence on the ``degree of disorder'' was recently observed for
several aperiodic perturbations \cite{berche97}.

The rest of the paper is organized as follows. In Sect.~2, we describe some
properties of the random lattices used in the simulations, and in Sect.~3 we
define the
model and give a few simulation details including estimates of autocorrelation
times. The quantities measured are defined in Sect.~4, where also their
theoretically expected FSS behavior is recalled.
In Sect.~5, we present the FSS
analysis of our data close to the transition point which yields
estimates for the critical exponents.
Finally, in Sect.~6 we present a discussion of our main results and close with
a few concluding remarks.

     \section{Random lattices} \label{sec:ranlat}
%
\paragraph{A. Voronoi/Delaunay random lattices:}
The notion of ``a random lattice'' is by no means unique and, in fact,
many different types of random lattices have been considered in the recent
literature \cite{dla,bhj94,jj98,bjj96,caer}. In this paper we concentrate
on so-called Poissonian Voronoi/Delaunay random lattices which, in arbitrary
dimensions, are defined as follows \cite{cfl,qft,ren}:
\begin{enumerate}

\item Draw $N$ sites $x_i$ at random in a unit volume (square in 2D,
      cube in 3D, \dots).
\vspace*{-0.14cm}
\item Associate with each site $x_i$ a Voronoi cell,
      $c_i \, \equiv \, \{x \,|\, d(x,x_i) \le d(x,x_j) \\\,\,\,\,\,\forall j \ne i \}$,
      which consists of all points $x$ that are closer to the center site $x_i$
      than to any other site. Here $d(x,y)$ denotes the usual Euclidean
      distance. This yields an irregular tessellation of the unit volume with
      D-dimensional Voronoi cells (polygons in 2D, polyhedra in 3D, \dots).
\vspace*{-0.14cm}
\item Construct the dual Delaunay lattice by linking the center sites of
      all Voronoi cells which share a common face.
\end{enumerate}
The first step approximates the Poissonian process. In the second step
we always
assume periodic boundary conditions, i.e., construct lattices of toroidal
topology. Using this construction the number of nearest neighbors of the
Delaunay lattice, the {\em local} coordination number $q$, varies randomly
from site to site. This constitutes the special type of quenched, random
connectivity disorder we are investigating in this work.

\begin{figure}[bhp]
{\Large \bf Figures}\\[2.0cm]
\vskip 6.8truecm
\includegraphics{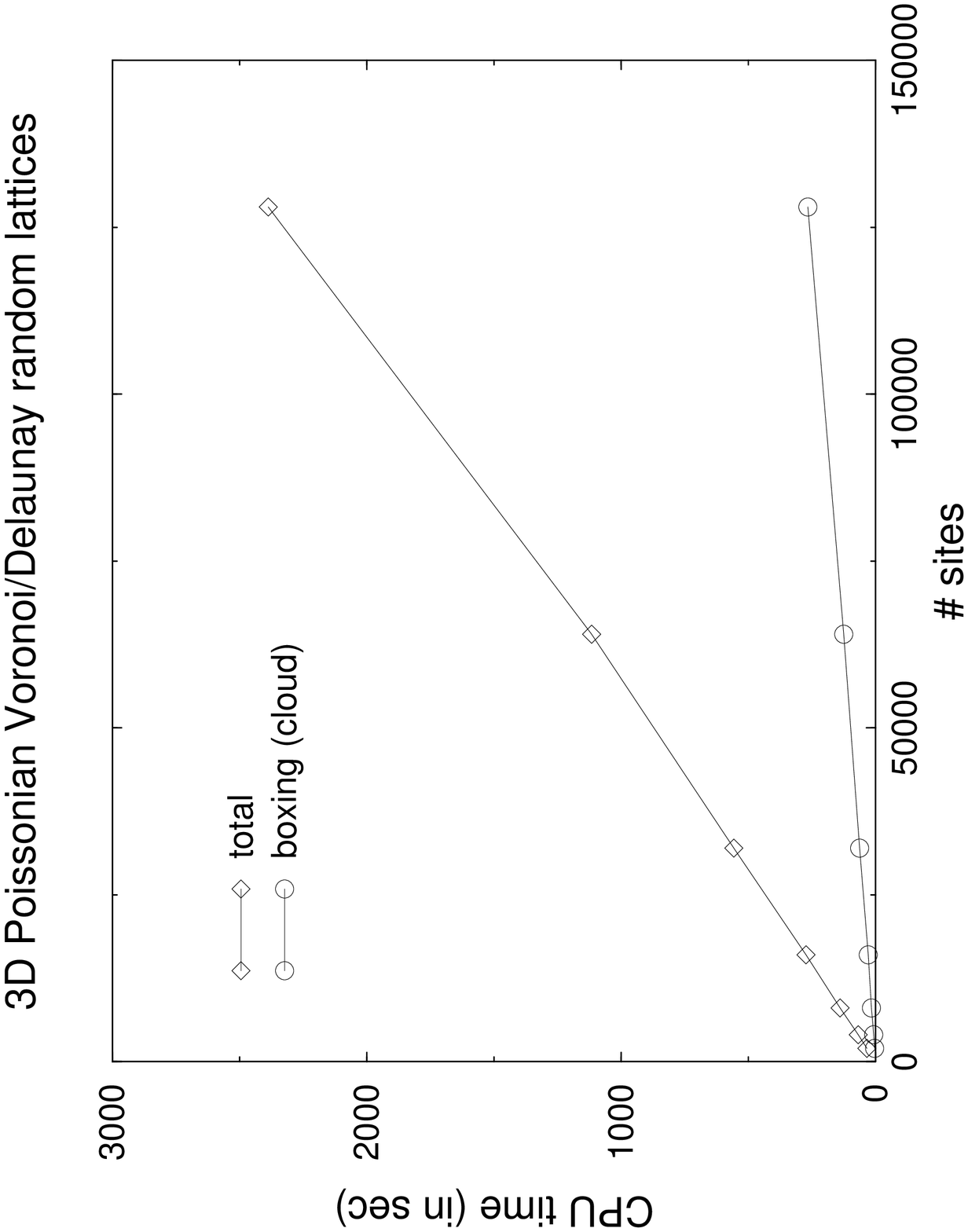}
  \caption[a]{{\em  The total CPU time spent in constructing a
  three-dimensional Poissonian random lattice according to the Voronoi/Delaunay
  prescription versus the number of sites $N$. The circles show the fraction
  of time needed to set up the ``cloud-of-neighbors''.}}
 \label{fig:latt_constr}
\end{figure}

The actually employed construction of the random lattices follows
loosely the method of Ref.~\cite{Tanemura} and is described in
the Appendix.
Following this method we succeeded to
reduce the complexity of the lattice construction for all practical purposes
from order $N^2$, as expected for the most straightforward implementation,
to order $N$, up to a small overhead $\propto N^2$ resulting from the initial
calculation of the distances between all pairs of two sites. The actually
measured CPU times as a function of $N$ are shown in
Fig.~\ref{fig:latt_constr}.
\paragraph{B. Random lattice properties:}
To test our random lattice construction we measured
several quantities which characterize the topology and geometry of the
lattices. These quantities are exactly known in the limit
$N \rightarrow \infty$. ``Topological'' quantities are the number
of links $N_l$, the number of triangles $N_{\triangle}$, and the
number of tetrahedra $N_{\tau}$, which according to Euler's theorem
should satisfy for a torus topology in any number of dimensions
and for {\em any} number of sites $N$ the relation
\begin{equation}
      N - N_l + N_{\triangle} - N_{\tau} = 0.
\label{eq:euler}
\end{equation}
We have of course explicitly tested that this topological constraint is
satisfied for all realizations, which is quite a sensitive confirmation
that the lattice construction works properly.
The measured averages
of $q$, $N_l$, $N_{\triangle}$, and $N_{\tau}$
over the 96 realizations used in the simulations
are collected in Table~\ref{tab:latt_stat1}.
The error bars are computed from the fluctuations among the 96 realizations.
We see that the
analytically known $N \rightarrow \infty$ limits are
approached very rapidly. The only exception is perhaps our
smallest lattice with $N = 2\,000$ sites
where deviations from the infinite-volume limit are clearly visible.
For $N \ge 4\,000$ in particular the average
number of nearest neighbors, $\overline{q}$, is fully consistent with the
theoretical value of
\begin{equation}
\overline{q} = 2 + \frac{48}{35} \pi^2 = 15.5354\dots .
\label{eq:q_theory}
\end{equation}
\begin{table}[tbp]
\centering
\caption[{\em Nothing.}]
 {{\em The average coordination number $\overline{q}$ and the total number of
 simplices normalized by the number of sites. The error bars are computed
 from the fluctuations among the 96 realizations.}}
\vspace{3ex}
\begin{tabular}{|r|llll|}
\hline
\hline
 & & & & \\[-0.4cm]
\multicolumn{1}{|c|}{$N$}          &
\multicolumn{1}{c}{$\overline{q}$}           &
\multicolumn{1}{c}{$N_l/N$}           &
\multicolumn{1}{c}{$N_{\triangle}/N$} &
\multicolumn{1}{c|}{$N_{\tau}/N$}      \\[0.1cm]
\hline
  2\,000 & 15.544(4)     & 7.7721(19)   & 13.5441(37)   & 6.7721(19)    \\
  4\,000 & 15.532(3)     & 7.7661(16)   & 13.5321(32)   & 6.7661(16)    \\
  8\,000 & 15.537(2)     & 7.7683(10)   & 13.5366(21)   & 6.7683(10)    \\
 16\,000 & 15.535(2)     & 7.76744(76)  & 13.5348(15)   & 6.76744(76)   \\
 32\,000 & 15.534(1)     & 7.76706(55)  & 13.5342(11)   & 6.76706(55)   \\
 64\,000 & 15.5351(7)    & 7.76755(36)  & 13.53507(72)  & 6.76755(36)   \\
128\,000 & 15.5349(5)    & 7.76743(26)  & 13.53486(53)  & 6.76743(26)   \\
exact    & 15.5354\dots  & 7.76772\dots & 13.53545\dots & 6.76772\dots \\
exact    & $2+\frac{48}{35}\pi^2$ & $1+\frac{24}{35}\pi^2$ &
              $\frac{48}{35}\pi^2$ & $\frac{24}{35}\pi^2$ \\[0.2cm]
\hline
\hline
\end{tabular}
\label{tab:latt_stat1}
\end{table}

As ``geometric'' quantities we measured the average volumes of the
simplices, i.e., the average link length $\overline{l} = \sum_{i=1}^{N_{\tau}}
\left( (\sum_{i \in \tau} l_i)/6 \right)/N_{\tau}$,
the average triangle area
$\overline{\triangle} = (\sum_{i=1}^{N_{\triangle}}
\triangle_i)/N_{\triangle}$,
and the average volume of a tetrahedron
$\overline{\tau} = (\sum_{i=1}^{N_{\tau}} \tau_i)/N_{\tau}$. Notice that the
average link length is defined in such a way that we first average over the
sides of a given tetrahedron and then over all
the tetrahedra. This is thus the average link length per
tetrahedron, and {\em not} the mean link length averaged over the whole
lattice, $\overline{l}^{(N)} = (\sum_{i=1}^{N_l} l_i)/N_l$, which is always
larger due to the fluctuations in the number of tetrahedra from realization
to realization. Our results normalized with an appropriate power of the
density $\rho = N/V$ are displayed in Table~\ref{tab:latt_stat2}.
From $N=4\,000$ on the numerical results are again fully consistent
with the analytical predictions as
$N \rightarrow \infty$.
Since all numbers agree very well with the analytical predictions we can be
quite sure that our lattice construction works properly and that we have
picked a typical sample of random lattices.

\begin{table}[htbp]
\centering
\caption[{\em Nothing.}]
 {{\em Average simplex volumes properly normalized to natural units.}}
\vspace{3ex}
\begin{tabular}{|r|llll|}
\hline
\hline
 & & & & \\[-0.4cm]
\multicolumn{1}{|c|}{$N$}          &
\multicolumn{1}{c}{$\overline{l}^{(N)}/\rho^{-1/3}$}           &
\multicolumn{1}{c}{$\overline{l}/\rho^{-1/3}$} &
\multicolumn{1}{c}{$\overline{\triangle}/\rho^{-2/3}$}     &
\multicolumn{1}{c|}{$\overline{\tau}/\rho^{-1}$}      \\[0.1cm]
\hline
  2\,000 & 1.28540(22)  & 1.23670(17)  & 0.597004(95)  & 0.147666(41)    \\
  4\,000 & 1.28566(18)  & 1.23717(13)  & 0.597362(70)  & 0.147797(34)    \\
  8\,000 & 1.28569(13)  & 1.23711(10)  & 0.597312(55)  & 0.147747(23)    \\
 16\,000 & 1.285623(75) & 1.237091(55) & 0.597306(31)  & 0.147766(17)    \\
 32\,000 & 1.285554(68) & 1.237027(49) & 0.597289(24)  & 0.147775(12)    \\
 64\,000 & 1.285537(40) & 1.237002(32) & 0.597272(19)  & 0.1477630(78)   \\
128\,000 & 1.285541(28) & 1.237012(20) & 0.597275(11)  & 0.1477666(58)   \\
exact    &              & 1.237033\dots& 0.597286\dots & 0.1477600\dots \\
exact    &  & $(\frac{3}{4 \pi})^{1/3}\frac{5 \cdot 7^3}{3^2 \cdot 4^4}
     \Gamma(\frac{1}{3})$   &
   $(\frac{3}{4 \pi})^{2/3}\frac{5^3 \cdot 7}{3^5 \pi}
     \Gamma(\frac{2}{3})$   &
   $\frac{35}{24 \pi^2}$ \\[0.2cm]
\hline
\hline
\end{tabular}
\label{tab:latt_stat2}
\end{table}

\begin{figure}[bhp]
\vskip 6.8truecm
\includegraphics{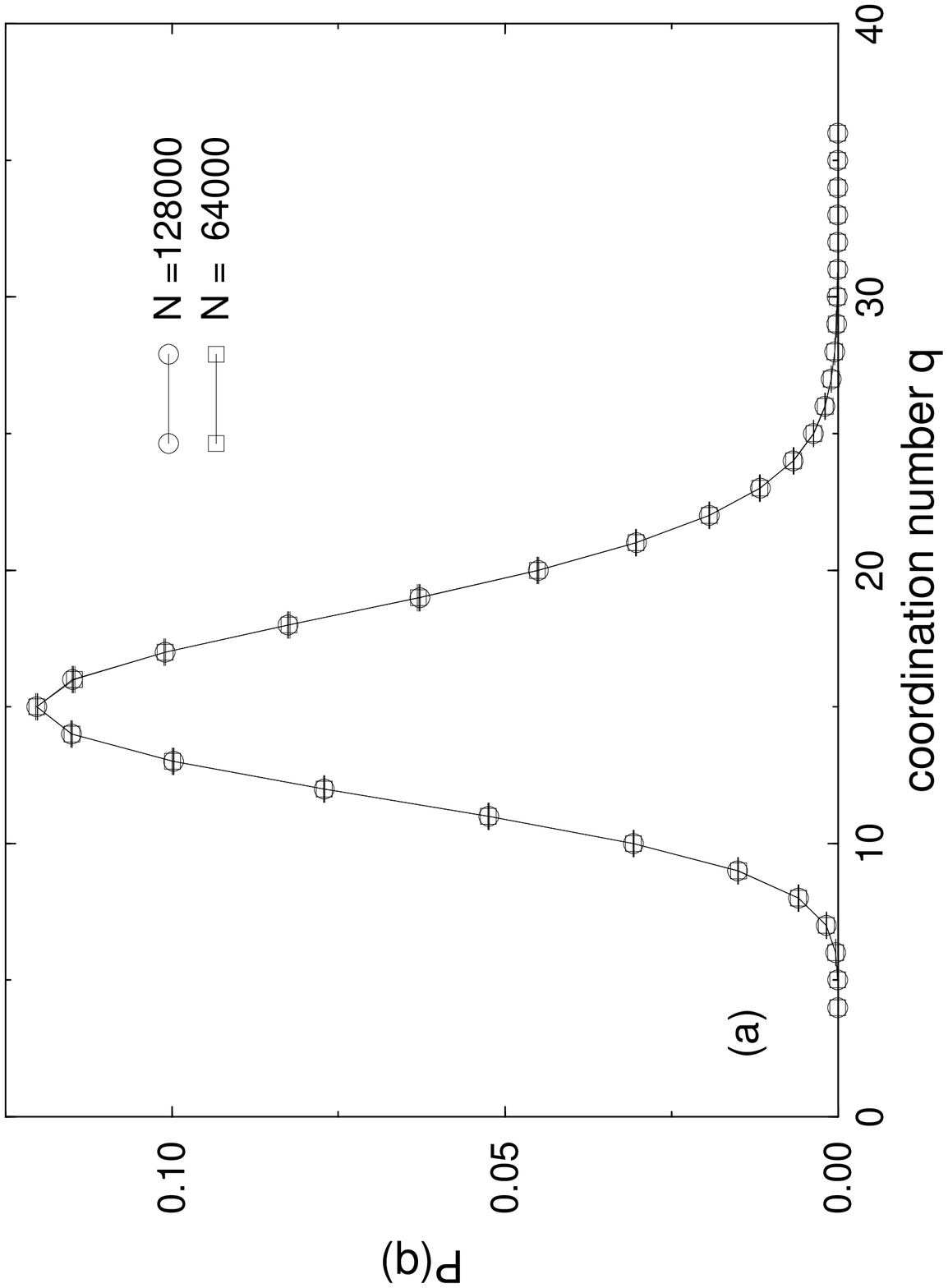}
\vskip 7.5truecm
\includegraphics{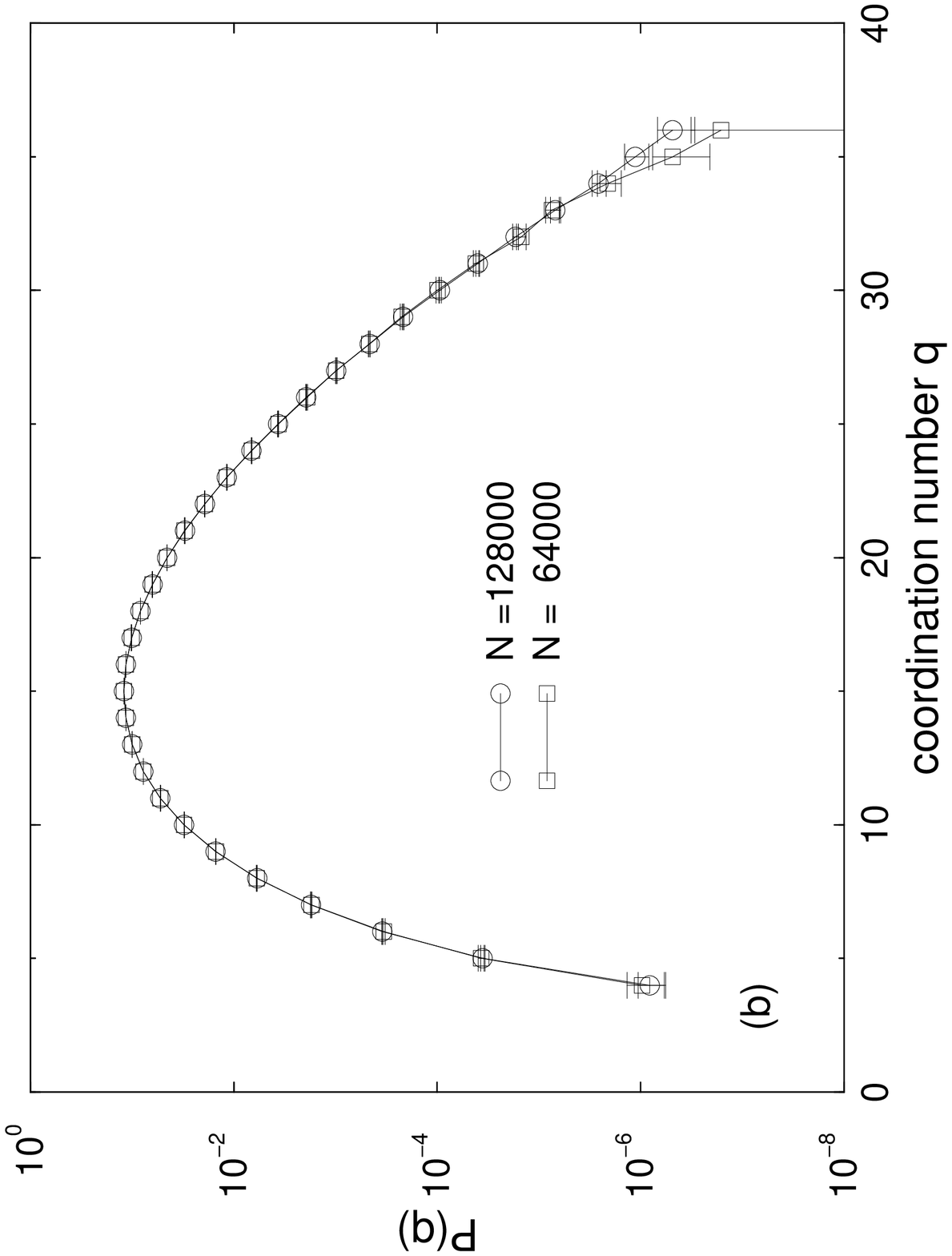}
\caption[a]{{\em (a) The probability density $P(q)$ of the coordination
numbers $q$.
The average is $\overline{q} = 2 + 48 \pi^2/35 = 15.5354\dots$.
(b) The same data as in (a) on a logarithmic scale.
        }}
\label{fig:P(q)}
\end{figure}

The (normalized) probability densities  $P(q)$ of the coordination numbers $q$
are shown in Fig.~\ref{fig:P(q)} for the lattices with $N = 64\,000$ and
$N=128\,000$ sites.
The average coordination number
is $\overline{q} =  2 + \frac{48}{35}\pi^2 = 15.5354\dots$. Due to the long
tail of $P(q)$ for large values of $q$, the maximum of $P(q)$ is found
at the next smaller integer number $q=15$, which occurs with a probability
of $12.03\%$.
%
     \section{The model and simulation techniques} \label{sec:model}
%
\paragraph{A. Model:}
We simulated the standard partition function of the Ising model,
\begin{equation}
Z = \sum_{\{s_i\}} e^{-K E};\mbox{~~} E = -\sum_{\langle ij \rangle} s_i s_j;
\mbox{~~} s_i=\pm 1,
\label{eq:1}
\end{equation}
where $K=J/k_BT > 0$ is the inverse temperature in natural units,
and $\langle ij \rangle$ denotes the
nearest-neighbor links of our three-dimensional toroidal random lattices.
In (\ref{eq:1}) we have adopted the convention used
in Ref.~\cite{jkv93}
and assigned to each link the same weight.

Another interesting option would be to assign to each link a weight
depending on the geometrical properties of the Voronoi/Delaunay construction
such as the length of the links or suitably defined areas of the associated
tesselation. In addition to the connectivity disorder this would introduce
also a (correlated) disorder of random-bond type. In order not to mix up these
two quenched disorder types, we decided to concentrate in this study
exclusively on the effect of the connectivity disorder.

\paragraph{B. Simulation:}
The finite-size scaling (FSS) analysis is performed on the basis of seven
different lattice sizes with $N=2\,000$, $4\,000$, $8\,000$, $16\,000$,
$32\,000$, $64\,000$, and $128\,000$ sites. For later use
we adopt the notation for regular lattices and define a linear lattice size
$L$ by $L=N^{1/3}$. The linear sizes of the lattices thus vary from
$L\approx 12.6$ to $L\approx 50.4$.
To investigate the dependence of thermal averages on
different realizations we performed MC simulations,
for each lattice size, on
the 96 randomly chosen
random lattice realizations discussed in Sect.~2.

For the update of the Ising spins
we employed Wolff's single-cluster (1C) algorithm \cite{1c}.
Various tests, in particular for the Ising model on two- and
three-dimensional regular lattices, have clearly demonstrated that
critical slowing down can be significantly reduced with this non-local
update algorithm \cite{uw_is,baillie,csd_1c}.
These tests also showed that in particular in three dimensions
the single-cluster algorithm is more efficient than the original
multiple-cluster formulation of Swendsen and Wang \cite{sw}.

All runs were performed in the vicinity of the critical point $K_c$. As a
first rough guess of $K_c$ we used the mean-field bound
$K_c \ge K_{\rm MF} = 1/\overline{q} \approx 0.064$. Due to the large average
coordination
number of 3D random lattices the mean-field approximation should be
better than for the simple cubic (SC) lattice where
$K_{\rm MF}^{\rm (SC)} = 1/6$ is about $1.33$ times
smaller than the actual $K_c^{\rm (SC)} \approx 0.221\,654\,6$.
Therefore, by applying the same correction factor to the random
lattice mean-field estimate we expect that $K_c$ is bounded from above
by $K_c \le 0.085$. This heuristic argument thus suggests that
$0.064 \le K_c \le 0.085$, such that $K_c \approx 0.075$ should
be a reasonable {\em a priori\/} guess.
Once good simulation points $K_0$ were estimated for the two smallest lattices
with $N=2\,000$ and $4\,000$ sites by determining for a few realizations the
location of the specific-heat and susceptibility maxima, we used
FSS extrapolations (assuming $\nu \approx 0.63$)
to predict $K_0$ for the larger lattices.

Depending on the lattice size from 30\,000 to 180\,000 clusters were
discarded to reach equilibrium from an initially completely ordered state.
Primary observables are the energy per spin, $e=E/N$, and the magnetization
per spin, $m = M/N = \sum_i s_i/N$, which were measured every $n_{\rm flip}$
cluster flip
and recorded in a time-series file. The average cluster size
$\langle |C| \rangle$ is an estimator for the reduced susceptibility in the
high-temperature phase, $\chi_{\rm red} = N \langle m^2 \rangle$, and
therefore scales with the lattice size according to $L^{\gamma/\nu}$,
where $\gamma$ and $\nu$ are the usual critical exponents. Since in three
dimensions $\gamma/\nu = 2 -\eta \approx 2$, we have to perform
$n_{\rm flip} \propto L$ cluster flips in order to flip, on the average, approximately
the same fraction of the total number of spins, $N = L^3$, for all lattice
sizes.
By adjusting the absolute scale of $n_{\rm flip}$ we made sure that for all
lattice
sizes the measurements were taken after about $N/2$ spin flips.
Since for the single-cluster update algorithm the autocorrelation times scale
only weakly with $L$ (see the discussion below) one then expects that, with
about the same number of measurements, the statistical
accuracy is comparable for all lattice sizes. With this set-up we performed
$100\,000$ measurements for each lattice size and realization ($110\,000$ for
$N=8\,000$). For more details on the employed statistics, see
Table~\ref{tab:stat}.
%
%
%
\begin{table}[tbp]
\centering
\caption[{\em Nothing.}]
{{\em Monte Carlo parameters of the simulations. $N \equiv L^3$ is the
lattice size, the hash mark symbol $\#$ denotes the number of realizations,
$K_0$ the inverse simulation temperature, $n_{\rm therm}$
the number of cluster flips during equilibration, and $n_{\rm meas}$ the
number of measurements taken every $n_{\rm flip}$ cluster flip.}}
\vspace{3ex}
\begin{tabular}{|r|rclrrr|}
\hline
\hline
 & & & & & & \\[-0.4cm]
\multicolumn{1}{|c|}{$N$}         &
\multicolumn{1}{c}{$L$}          &
\multicolumn{1}{c}{\#}  &
\multicolumn{1}{c}{$K_0$}    &
\multicolumn{1}{c}{$n_{\rm therm}$}       &
\multicolumn{1}{c}{$n_{\rm meas}$}       &
\multicolumn{1}{c|}{$n_{\rm flip}$}          \\[0.1cm]
\hline
   2\,000 & 12.6 & 96 & 0.0735    &  30\,000  & 100\,000  &  4 \\
   4\,000 & 15.9 & 96 & 0.0735    &  50\,000  & 100\,000  &  5 \\
   8\,000 & 20.0 & 96 & 0.0732    &  66\,000  & 110\,000  &  6 \\
  16\,000 & 25.2 & 96 & 0.0729    &  80\,000  & 100\,000  &  8 \\
  32\,000 & 31.7 & 96 & 0.0728\,5 & 100\,000  & 100\,000  & 10 \\
  64\,000 & 39.1 & 96 & 0.0726\,5 & 150\,000  & 100\,000  & 13 \\
 128\,000 & 50.4 & 32 & 0.0725\,3 & 180\,000  & 100\,000  & 16 \\
 128\,000 & 50.4 & 64 & 0.0725\,9 & 180\,000  & 100\,000  & 16 \\[0.1cm]
\hline
\hline
\end{tabular}
\label{tab:stat}
\end{table}
%
\paragraph{C. Update dynamics:}
A useful measure of the update dynamics is the integrated autocorrelation
time $\hat{\tau}$ \cite{tau_reviews}.
To estimate $\hat{\tau}$ for the measurements of the energy $e$
and the magnetization $m$
we used
the fact that $\hat{\tau}$ enters in
the error estimate $\epsilon^2 = \sigma^2 2\hat{\tau}/n_{\rm meas}$ for the
mean value $\overline{O}$ of $n_{\rm meas}$ correlated measurements with
variance
$\sigma^2 = \langle O; O \rangle \equiv
\langle O^2 \rangle - \langle O \rangle^2$, and determined $\epsilon^2$
by blocking procedures. Using 100 blocks of 1000 measurements each
we obtained the results
compiled in Table~\ref{tab:tau} where all results are averaged over the
96 randomly chosen realizations. We see that the integrated autocorrelation
times for the measurements of $e$ and $m$
are of the order of $\hat{\tau}_e \approx 2.5 - 3.5$ and
$\hat{\tau}_{m} \approx 2.4 - 3.0$. Since completely uncorrelated data
correspond to $\hat{\tau}=0.5$, our thermal sample thus effectively consists
of about $15\,000-20\,000$ uncorrelated measurements for each of the 96
realizations. This amounts to a total of $(1.5 - 2.0)\,\,\! \times 10^6$
effectively uncorrelated
measurements for each lattice size.
%
%
%
%
\begin{table}[tp]
\scriptsize
\centering
\caption[{\em Nothing.}]
{{\em Average cluster size $[\langle|C| \rangle ]_{\rm av}$ and
autocorrelation times of energy and magnetization at the simulation point
$K_0$, where $\tau = f \cdot \hat{\tau}$ and
$f = n_{\rm flip} \langle |C| \rangle /N$.
The $tau$'s are obtained from a blocking analysis on the basis of
100 block.}}
\vspace{3ex}
\begin{tabular}{|r|llrcrrc|}
\hline
\hline
 & & & & & & & \\[-0.2cm]
\multicolumn{1}{|c|}{$N$}          &
\multicolumn{1}{c}{$[\langle|C| \rangle ]_{\rm av}$}         &
\multicolumn{1}{c}{$[\hat{\tau}_e]_{\rm av}$}       &
\multicolumn{1}{c}{$[\tau_e]_{\rm av}$}             &
\multicolumn{1}{c}{$[f]_{\rm av} \cdot [\hat{\tau}_e]_{\rm av}$}       &
\multicolumn{1}{c}{$[\hat{\tau}_m]_{\rm av}$}       &
\multicolumn{1}{c}{$[\tau_m]_{\rm av}$}             &
\multicolumn{1}{c|}{$[f]_{\rm av} \cdot [\hat{\tau}_m]_{\rm av}$} \\[0.15cm]
\hline
   2\,000 & \,~246.9(1.9)
      & 2.529(40)
      & 1.245(19)
      & 1.249
      & 2.434(38)
      & 1.200(20)
      & 1.202                       \\
   4\,000 & \,~453.3(3.9)
      & 2.443(42)
      & 1.375(20)
      & 1.384
      & 2.412(38)
      & 1.360(20)
      & 1.367                       \\
   8\,000 & \,~740.1(5.5)
      & 2.702(42)
      & 1.494(22)
      & 1.500
      & 2.594(40)
      & 1.436(22)
      & 1.440                       \\
  16\,000 & 1084.0(9.5)
      & 2.955(43)
      & 1.593(21)
      & 1.602
      & 2.730(38)
      & 1.475(22)
      & 1.480                        \\
  32\,000 & 1962(17)
      & 2.741(50)
      & 1.666(24)
      & 1.681
      & 2.572(44)
      & 1.567(23)
      & 1.577                        \\
  64\,000 & 2678(24)
      & 3.177(55)
      & 1.718(27)
      & 1.728
      & 2.795(42)
      & 1.516(23)
      & 1.520                        \\
 128\,000 & 3521(87)
      & 3.92(11)
      & 1.705(42)
      & 1.725
      & 3.157(72)
      & 1.385(42)
      & 1.390                        \\
 128\,000 & 4325(41)
      & 3.399(72)
      & 1.824(31)
      & 1.838
      & 2.964(60)
      & 1.593(29)
      & 1.602                        \\[0.1cm]
\hline
\hline
\end{tabular}
\label{tab:tau}
\normalsize
\end{table}

While this properly characterizes the effective statistics of our simulations,
the numbers for $\hat{\tau}$ of a single-cluster simulation are not yet well
suited for a comparison with other update algorithms or with single-cluster
simulations on regular lattices.
To get a comparative work-estimate, the usual procedure \cite{1c}
is to convert the $\hat{\tau}$ by multiplying with a factor
$f = n_{\rm flip} \langle |C| \rangle/N$ to a scale where, on the average,
measurements are taken after every spin has been flipped once (similar to,
e.g., Metropolis simulations). For quenched, random systems this procedure
is not unique due to the necessary average over realizations. In
Table~\ref{tab:tau} we therefore present both options, $[\tau]_{\rm av} \equiv
[f \cdot \hat{\tau}]_{\rm av}$ and also
$[f]_{\rm av} \cdot [\hat{\tau}]_{\rm av}$. The differences between the
two averaging prescriptions are, however, extremely small.

The numbers in Table~\ref{tab:tau} obtained in this way are very similar
to results for
the regular simple cubic (SC) lattice \cite{uw_is,baillie}. By fitting
a power law, $\tau \propto L^z$, to the data for the five largest lattices
we obtain $[\tau_e]_{\rm av} = 0.82(6) L^{0.20(2)}$, and
$[\tau_{m}]_{\rm av} = 1.07(8) L^{0.10(2)}$, respectively. The dynamic critical
exponents $z$ for the
random lattice simulations should be compared with $z_e = 0.28(2)$ and
$z_{\chi} = 0.14(2)$ for the SC lattice \cite{uw_is}.

%
\section{Observables and finite-size scaling}  \label{sec:obs}
%
From the time series of $e$ and $m$ it is straightforward to
compute in the FSS region various quantities at nearby values of $K_0$ by
standard reweighting methods \cite{fs1}.
For the estimation of the statistical (thermal) errors,
for each of the 96 realizations the time-series data was split into
100 bins, which were jack-knifed \cite{jack} to decrease the bias in the
analysis of reweighted data.
The final values are averages over the 96 realizations which will be denoted
by square brackets $[\dots]_{\rm av}$, and the error bars are computed from the
fluctuations among the realizations. Note that these errors contain both,
the average thermal error for a given realization and the theoretical variance
for infinitely accurate thermal averages which is caused by the variation of
the quenched, random geometry of the 96 lattices.

From the time series of the energy measurements we can compute by reweighting
the average energy, specific heat, and energetic fourth-order parameter,
\begin{eqnarray}
u(K) &=& [\langle E \rangle]_{\rm av}/N,\\
C(K) &=& K^2 \, N [\langle e^2 \rangle - \langle e \rangle^2]_{\rm av},\\
V(K) &=&  [1 - \frac{\langle e^4 \rangle}{3 \langle e^2 \rangle^2}]_{\rm av}.
\label{eq:ene}
\end{eqnarray}
Similarly we can derive from the magnetization measurements the average
magnetization, the susceptibility, and the magnetic cumulants,
\begin{eqnarray}
m(K) &=& [\langle |m| \rangle]_{\rm av},\\
\chi(K) &=& K \, N [ \langle m^2 \rangle - \langle |m| \rangle^2 ]_{\rm av},\\
U_2(K) &=& [1 - \frac{\langle m^2 \rangle}{3 \langle |m|
\rangle^2}]_{\rm av},\\
U_4(K) &=& [1 - \frac{\langle m^4 \rangle}{3 \langle m^2 \rangle^2}]_{\rm av}.
\label{eq:mag}
\end{eqnarray}
Further useful quantities involving both the energy and magnetization are
\begin{eqnarray}
\frac{d [\langle |m| \rangle]_{\rm av}} {d K} &=&
  [\langle |m| E \rangle  - \langle |m| \rangle \langle E \rangle]_{\rm av},\\
\frac{d \ln [\langle |m| \rangle]_{\rm av}} {d K} &=&
\left[ \frac{\langle |m| E \rangle}{\langle |m| \rangle} - \langle E
\rangle\right]_{\rm av},\\
\frac{d \ln [\langle m^2 \rangle]_{\rm av}} {d K} &=&
\left[ \frac{\langle m^2 E \rangle}{\langle m^2 \rangle} - \langle E
\rangle \right]_{\rm av}.
\end{eqnarray}

In the infinite-volume limit these quantities exhibit singularities at
the transition point. In finite systems the singularities are smeared
out and scale in the critical region according to
\begin{eqnarray}
C &=& C_{\rm reg} + L^{\alpha/\nu} f_C(x) [1 + \dots],
\label{eq:fss_C}\\
 {[} \langle |m| \rangle ]_{\rm av} &=&
L^{-\beta/\nu} f_{m}(x) [1 + \dots],
\label{eq:fss_m}\\
\chi &=& L^{\gamma/\nu} f_\chi(x) [1 + \dots],
\label{eq:fss_chi}\\
\frac{d [\langle |m| \rangle]_{\rm av}} {d K} &=&
L^{(1-\beta)/\nu} f_{m'}(x) [1 + \dots],
\label{eq:fss_dmdk}\\
\frac{d \ln [\langle |m|^p \rangle]_{\rm av}} {d K} &=&
L^{1/\nu} f_p(x) [1 + \dots],
\label{eq:fss_dlnmdk}\\
\frac{d U_{2p}}{d K} &=& L^{1/\nu} f_{U_{2p}}(x) [1 + \dots],
\label{eq:fss_dUdk}
\end{eqnarray}
where $C_{\rm reg}$ is a regular background term, $\nu$, $\alpha$,
$\beta$, and $\gamma$, are the usual critical exponents,
$f_i(x)$ are FSS functions with
\begin{equation}
x = (K - K_c) L^{1/\nu}
\end{equation}
being the scaling variable, and the brackets $[1 + \dots]$ indicate
corrections-to-scaling terms which become unimportant for sufficiently
large system sizes $L$.
%
     \section{Results}    \label{sec:results}
\begin{figure}[bhp]
\vskip 7.8truecm
\includegraphics{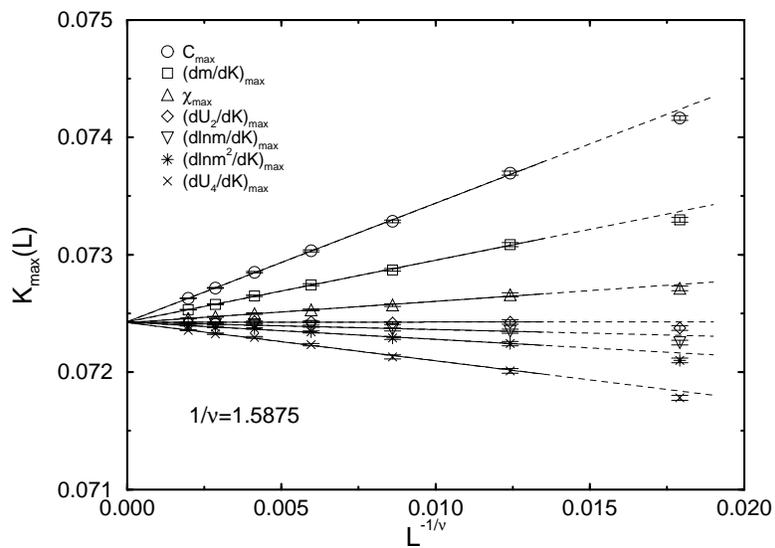}
\caption[a]{{\em FSS fits of the pseudo-transition points
$K_{{\rm max}_i}$ with $1/\nu = 1.5875$ fixed, yielding a combined estimate of
$K_c = 0.0724\,249(40)$.
        }}
\label{fig:kc}
\end{figure}

%
By applying standard reweighting techniques to each of the 96 time-series data
we first determined the temperature dependence of $C_i(K)$, $\chi_i(K)$,
\dots, $i=1,\dots,96$, in the neighborhood of the simulation point $K_0$. By
estimating the valid reweighting range for each of the realizations we made
sure that no systematic errors crept in by this procedure. Once the
temperature dependence is known for each realization, we can easily compute
the disorder average, e.g.,
$C(K) = \sum_{i=1}^{96} C_i(K)/96$,
and then determine the maxima of the averaged quantities, e.g.,
$C_{\rm max}(K_{{\rm max}_C}) \equiv \max_K C(K)$. The
locations of the maxima of $C$, $\chi$, $dU_2/dK$, $dU_4/dK$,
$d[\langle |m| \rangle]_{\rm av}/dK$,
$d \ln [ \langle |m| \rangle]_{\rm av}/dK$, and
$d \ln [ \langle m^2 \rangle]_{\rm av}/dK$ provide us with seven sequences
of pseudo-transition points $K_{{\rm max}_i}(L)$ which all should scale
according to $K_{{\rm max}_i}(L) = K_c + a_i L^{-1/\nu} + \dots$.
In other words, the scaling variable
$x = (K_{{\rm max}_i}(L) - K_c) L^{1/\nu} = a_i + \dots$ should be
constant, if we neglect the small higher-order corrections indicated by
$\dots$. To give an idea on how these sequences approach $K_c$ we show below in
Table~\ref{tab:kc} besides the estimates for $K_c$ also the values of
$a_i, i=1,\dots,7$ (see also
Fig.~\ref{fig:kc}).

It should be emphasized that while the precise
estimates of $a_i$ do depend on the value of $\nu$, the qualitative
conclusion that $x \approx {\rm const.}$ for $K_{{\rm max}_i}$
does not require any {\em a priori\/} knowledge of $\nu$ or $K_c$.
Using this information we have thus several possibilities to
extract unbiased estimates of the critical exponents $\nu$, $\alpha/\nu$,
$\beta/\nu$, and $\gamma/\nu$ from least-squares fits assuming
the FSS behaviors
(\ref{eq:fss_C})-(\ref{eq:fss_dUdk}). Once $\nu$ is estimated we can then
use $K_{{\rm max}_i}(L) = K_c + a_i L^{-1/\nu} + \dots$ to extract also
$K_c$ and $a_i$.
\paragraph{A. Critical exponent $\nu$:}
Let us thus start with the correlation length exponent $\nu$ for which we
can obtain
$4 \times 7 = 28$ different estimates by considering the scaling of
$d \ln [\langle |m| \rangle]_{\rm av}/dK$,
$d \ln [\langle m^2 \rangle]_{\rm av}/dK$,
$d U_2/d K$, and $d U_4/d K$ at the seven sequences of
pseudo-transition points $K_{{\rm max}_i}(L)$. Of course, these estimates
are statistically {\em not} uncorrelated, but they are differently
affected by corrections to the leading FSS behavior. To test the
importance of these corrections to scaling we first estimated $\nu$ from
fits using all available lattice sizes, and then compared with the results of
fits where the two smallest sizes were successively discarded. As a result we
did not observe any systematic improvement by omitting the smallest lattices.
In fact, already for the fits using all sizes, we obtained goodness-of-fit
parameters \cite{numlib} $Q \ge 0.3$ for about $80\%$ of the 28 fits. The only
unacceptable fit was that of $d \ln [\langle m^2 \rangle]_{\rm av}/dK$ at the
$K_{{\rm max}_\chi}$ sequence with $Q=0.01$. Here we omitted the $N=2000$
data point which improves the goodness to $Q=0.22$.

This analysis clearly
shows that with our data there is no need to include corrections-to-scaling
terms in the fits which would necessitate non-linear fitting procedures which
usually tend to be quite unstable. Of course, such a conclusion depends
on the accuracy of the data and only shows that the correction terms are so
small that they cannot be resolved within our accuracy. This is quite
remarkable because in the pure 3D Ising model study of Ref.~\cite{Ba_etal99}
for lattices of size $L = 8$ to $128$ (using a different FSS technique)
confluent corrections to scaling $\propto L^{-\omega}$ with $\omega = 0.87(9)$
could be resolved. One possible explanation is that random lattices with
large average coordination number and preserved rotational invariance indeed
reach the infinite-volume limit earlier than their regular counterparts. This
is quite conceivable since the magnitude of the correction term does not only
depend on the exponent but also on its amplitude, and the latter is a
non-universal quantity which does definitely depend on the lattice structure.
%
%
\begin{table}[htbp]
\centering
\caption[{\em Nothing.}]
 {{\em Fit results for the critical exponent $1/\nu$.}}
\vspace{3ex}
\begin{tabular}{|r|r|r|}
\hline\hline
& & \\[-0.4cm]
\multicolumn{1}{|c|}{quantity}        &
\multicolumn{1}{c|}{type}         &
\multicolumn{1}{c|}{$1/\nu$}      \\[0.1cm]
\hline
$d U_2/dK$ & at maximum   &  1.5818(33) \\
           & average      &  1.5783(32) \\
           & weighted av. &  1.5795(12) \\
           & final        &  1.5795(27) \\
\hline
$d U_4/dK$ & at maximum   &  1.5797(42) \\
           & average      &  1.5763(34) \\
           & weighted av. &  1.5774(13) \\
           & final        &  1.5774(25) \\
\hline
$dU_{2p}/dK$  & average      &  1.5773(23) \\
$p=1$ and 2& weighted av. &  1.5785(09) \\
           & final        &  1.5785(25) \\
\hline\hline
$d \ln [\langle |m| \rangle]_{\rm av}/dK$
           & at maximum   &  1.5886(16) \\
           & average      &  1.5888(08) \\
           & weighted av. &  1.5889(06) \\
           & final        &  1.5889(14) \\
\hline
$d \ln [\langle m^2 \rangle]_{\rm av}/dK$
           & at maximum   &  1.5894(15) \\
           & average      &  1.5902(07) \\
           & weighted av. &  1.5898(06) \\
           & final        &  1.5898(12) \\
\hline
$d \ln [\langle m^p \rangle]_{\rm av}/dK$
           & average      &  1.5895(06) \\
$p=1$ and 2& weighted av. &  1.5894(05) \\
           & final        &  1.5894(12) \\
\hline\hline
$dU_{2p}/dK_{\rm max}$ and
           & average      &  1.5849(25) \\
$d \ln [\langle m^p \rangle]_{\rm av}/dK_{\rm max}$
           & weighted av. &  1.5878(10) \\
(4 fits)
           & final        &  1.5878(15) \\
\hline\hline
$dU_{2p}/dK$ and & average      &  1.5834(16) \\
$d \ln [\langle m^p \rangle]_{\rm av}/dK$  & weighted av. &  1.5875(04) \\
(28 fits)           & final        &  1.5875(12) \\[0.1cm]
\hline\hline
\end{tabular}
\label{tab:nu_inv}
\end{table}

Having now 28 estimates of $1/\nu$ from
fits with $Q \ge 0.15$ we proceeded as follows. We computed arithmetic and
error weighted averages for the four subgroups of estimates and finally
over the total set of estimates. The main results are collected in
Table~\ref{tab:nu_inv}.
Because of the neglected cross-correlations
between the fit results, in particular the error estimate of the weighted
average should be taken with some care. As our best estimates we therefore
quote throughout this paper in all tables in the lines labeled ``final''
always the weighted mean and quite conservatively the smallest error among
all available fits (making thus the plausible assumption that the error of the
weighted average is never larger than the error of the most accurate
fit result that contributes to this average). As a general trend we see in
Table~\ref{tab:nu_inv} that the
fits of $d \ln [\langle |m|^p \rangle]_{\rm av}/dK$
yield more accurate results than those of
$dU_{2p}/dK$. We also observe, however, that
the partial averages over the fits of $dU_{2p}/dK, p=1,2$ and
$d \ln [\langle |m|^p \rangle]_{\rm av}/dK, p=1,2$, are only
marginally consistent, even though in all cases the goodness-of-fit was high.
We thus have no reason based on statistical arguments to favor one or the
other group of fits and therefore take
as our final value the weighted average over all 28 estimates which yields
\begin{equation}
1/\nu = 1.5875 \pm 0.0012, \qquad \nu = 0.62992 \pm 0.00048,
\label{eq:nu_all}
\end{equation}
with the
minimal error coming from the fit of
$d \ln [\langle |m|^2 \rangle]_{\rm av}/dK$
at the maximum locations of $d [\langle |m| \rangle]_{\rm av}/dK$.

\begin{figure}[t]
\vskip 4.7truecm
\includegraphics{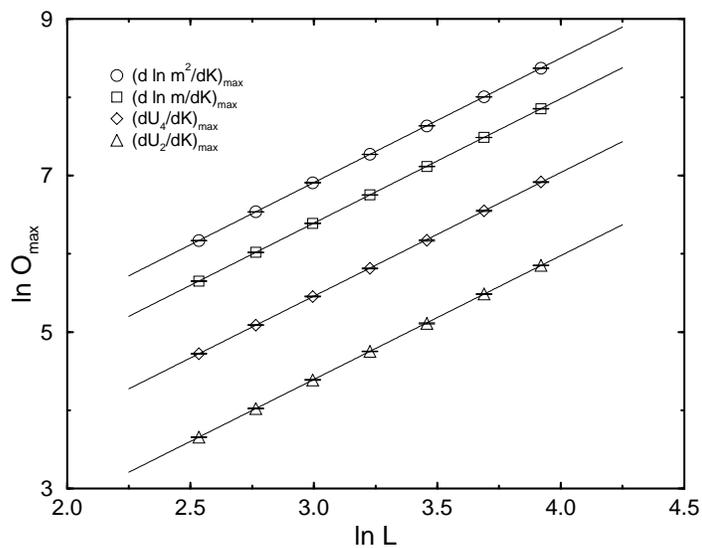}
\caption[a]{{\em FSS fits to extract $1/\nu$.
        }}
\label{fig:nufit}
\end{figure}

If we only average the results of the fits of the maxima of
$d \ln [\langle |m| \rangle]_{\rm av}/dK$,
$d \ln [\langle m^2 \rangle]_{\rm av}/dK$,
$d U_2/d K$, and $d U_4/d K$, we obtain basically the same final
estimate with a slightly larger error bar:
\begin{equation}
1/\nu = 1.5878 \pm 0.0015, \qquad \nu = 0.62980 \pm 0.00060.
\label{eq:nu_max}
\end{equation}
To give also a visual impression of the quality of these fits, they
are shown in Fig.~\ref{fig:nufit}. This reconfirms the
weighted average (\ref{eq:nu_all}) over all 28 fits, and
our final estimate for the correlation length exponent is thus
\begin{equation}
\nu = 0.6299 \pm 0.0005.
\label{eq:nu_final}
\end{equation}
For comparison, for Ising models on regular simple cubic (SC)
lattices Ferrenberg and Landau \cite{fl3dis} obtained $\nu =
0.6289(8)$, Bl\"ote {\em et al.} \cite{bloete1} concluded that
$\nu = 0.6301(8)$, Ballesteros {\em et al.} \cite{Ba_etal99} found
\cite{foot1_sytematic} $\nu = 0.6294(5)[5]$, and in a recent study
of the $\phi^4$ lattice field theory Hasenbusch \cite{hasenbusch}
estimated $\nu = 0.6296(3)[4]$. Within error bars all three
estimates are in perfect agreement with our result
(\ref{eq:nu_final}) for random lattices.

\paragraph{B. Critical coupling $K_c$:}
Having determined the critical exponent $\nu$, it is straightforward to
obtain estimates of the critical coupling $K_c$ from linear least-squares
fits to
\begin{equation}
K_{\rm max_i} = K_c + a_i L^{-1/\nu},
\end{equation}
where $K_{\rm max_i}$ are the seven pseudo-transition points discussed
earlier. Here we found a significant improvement of the quality of the fits
if the smallest lattice size with $N=2\,000$ was excluded. This can also be
inspected visually in Fig.~\ref{fig:kc}, where the data and fits are shown.
We see a systematic trend that the $N=2\,000$ data lie a little bit too low.
In Table~\ref{tab:kc} we therefore display the fit results over the six
lattice sizes $N=4\,000 - 128\,000$. By using the same averaging procedure as
before we arrive at the final estimate
\begin{equation}
K_c = 0.0724\,249 \pm 0.0000\,040.
\label{eq:kc}
\end{equation}
Of course, in principle this estimate is biased by our estimate of $\nu$.
We have checked, however, that the dependence on $\nu$ is extremely weak.
If we repeat the fits with $1/\nu = 1.5875 \pm 3 \epsilon_{1/\nu}$, where
$\epsilon_{1/\nu} = 0.0012$ is the error on the estimate of $1/\nu$ in
(\ref{eq:nu_all}), we obtain a variation in the estimate for $K_c$ by
only $\pm 2$ in the last digit, which is much smaller than the statistical
 error in (\ref{eq:kc}).

%
%
\begin{table}[t]
\centering
\caption[{\em Nothing.}]
 {{\em Fit results for the critical coupling $K_c$,
 using the FSS ansatz $K_{{\rm max}_i} = K_c + a_i L^{-1/\nu}$
 and our best estimate of $1/\nu = 1.5875(12)$. The fit range is
 always from $N=4\,000$ to $128\,000$.}}
\vspace{3ex}
\begin{tabular}{|r|lrl|}
\hline\hline
& & & \\[-0.4cm]
\multicolumn{1}{|c|}{$K_{\rm max}$ of}  &
\multicolumn{1}{c}{$K_c$}               &
\multicolumn{1}{c}{$a$}                 &
\multicolumn{1}{c|}{$Q$}                \\[0.1cm]
\hline
 $C$                           & 0.0724\,282(40) & $ 0.1010(12)$ & 0.63 \\
 $\chi$                        & 0.0724\,210(40) & $ 0.0182(11)$ & 0.87 \\
 $dU_4/dK$                     & 0.0724\,221(54) & $-0.0326(14)$ & 0.65 \\
 $dU_2/dK$                     & 0.0724\,238(44) & $ 0.0002(12)$ & 0.89 \\
 $d [\langle |m| \rangle]_{\rm av}/dK$ & 0.0724\,277(40) & $ 0.0525(11)$ & 0.82 \\
 $d\ln[\langle |m| \rangle]_{\rm av}/dK$  & 0.0724\,250(43) & $-0.0063(12)$
                                          & 0.83 \\
 $d\ln[\langle m^2 \rangle]_{\rm av}/dK$  & 0.0724\,253(44) & $-0.0146(12)$
                                          & 0.76 \\
 \hline
 average                       & 0.0724\,247(10) &            &      \\
 weighted av.                  & 0.0724\,249(16) &            &      \\
 final                         & 0.0724\,249(40) &            &      \\[0.1cm]
 \hline\hline
\end{tabular}
\label{tab:kc}
\end{table}

\paragraph{C. Critical exponent $\gamma$:}

The exponent ratio $\gamma/\nu$ can be obtained from fits to the FSS
behavior (\ref{eq:fss_chi}) of the susceptibility.
By monitoring the quality of the fits we decided to discard the $N=2\,000$
data for the $K_{{\rm max}_C}$ and $K_{{\rm max}_{dm/dK}}$ sequences (which
led to $Q$ values of $0.05$ and $0.02$, respectively). The fits collected
in Table~\ref{tab:other} then all have $Q \ge 0.25$. The final result is
\begin{equation}
\gamma/\nu = 1.9576 \pm 0.0013,
\label{eq:gamma_nu}
\end{equation}
which should be compared with the estimates for regular SC lattices of
$\gamma/\nu = 1.970(14)$ in Ref.~\cite{fl3dis},
$\gamma/\nu = 1.9630(30)$ in Ref.~\cite{bloete1},
$\gamma/\nu = 1.9626(6)[6]$ in Ref.~\cite{Ba_etal99}, and
$\gamma/\nu = 1.9642(4)[5]$ in Ref.~\cite{hasenbusch}.

For the exponent $\eta$, the estimate
(\ref{eq:gamma_nu}) implies
\begin{equation}
\eta = 2 - \gamma/\nu = 0.0424 \pm 0.0013,
\label{eq:eta}
\end{equation}
and, by using our value (\ref{eq:nu_all}) for $1/\nu$, we derive
\begin{equation}
\gamma = 1.2332 \pm 0.0018.
\label{eq:gamma}
\end{equation}

%
%
\begin{table}[t]
\scriptsize
\centering
\caption[{\em Nothing.}]
 {{\em Fit results for the critical exponents $\gamma/\nu$,
 $\beta/\nu$, and $(1-\beta)/\nu$. The superscripts $*$ and
 $\#$ at the $Q$ values indicate that these fits start at $N=4\,000$
 and $N=8\,000$, respectively. The other fits use all data
 from $N=2\,000$ to $128\,000$.}}
\vspace{3ex}
\begin{tabular}{|r|ll|ll|ll|}
\hline\hline
& & & & & & \\[-0.2cm]
\multicolumn{1}{|c|}{at $K_{\rm max}$ of}  &
\multicolumn{1}{c}{$\gamma/\nu$}           &
\multicolumn{1}{c|}{$Q$}                   &
\multicolumn{1}{c}{$\beta/\nu$}            &
\multicolumn{1}{c|}{$Q$}                   &
\multicolumn{1}{c}{$(1-\beta)/\nu$}        &
\multicolumn{1}{c|}{$Q$}                   \\[0.1cm]
\hline
 $C$                           & 1.9357(37)  & $0.49^*$ & 0.5089(11)  & 0.10     & 1.0690(32)  & $0.82^{\#}$ \\
 $\chi$                        & 1.9551(13)  & 0.36     & 0.52374(82) & $0.85^*$ & 1.0663(14)  & 0.30        \\
 $dU_4/dK$                     & 1.9641(25)  & 0.35     & 0.5180(41)  & 0.27     & 1.0707(39)  & 0.20        \\
 $dU_2/dK$                     & 1.9581(13)  & 0.59     & 0.5173(22)  & 0.31     & 1.0713(19)  & 0.15        \\
 $d [\langle |m| \rangle]_{\rm av}/dK$      & 1.9476(19)  & $0.29^*$ & 0.51097(95) & $0.10^*$ & 1.0683(24)  & $0.65^{\#}$ \\
 $d\ln[\langle |m| \rangle]_{\rm av}/dK$   & 1.9603(13)  & 0.76     & 0.5140(19)  & 0.28     & 1.0657(29)  & $0.90^{\#}$ \\
 $d\ln[\langle m^2 \rangle]_{\rm av}/dK$ & 1.9626(13)  & 0.77     & 0.5128(22)  & 0.30     & 1.0664(30)  & $0.69^{\#}$ \\
 \hline
 average                       & 1.9548(38)  &        & 0.5151(19)  &        & 1.06822(84) &           \\
 weighted av.                  & 1.95758(59) &        & 0.51587(49) &        & 1.06798(86) &           \\
 final                         & 1.9576(13)  &        & 0.51587(82) &        & 1.0680(14)  &           \\[0.1cm]
 \hline\hline
\end{tabular}
\label{tab:other}
\normalsize
\end{table}

\paragraph{D. Critical exponent $\beta$:}

The exponent ratio $\beta/\nu$ can be either obtained from the FSS behavior
of $[\langle |m| \rangle]_{\rm av}$ or $d [\langle |m| \rangle]_{\rm av}/dK$,
Eqs.~(\ref{eq:fss_m}) or (\ref{eq:fss_dmdk}). In the first case, the
sequences $K_{{\rm max}_\chi}$ and $K_{{\rm max}_{dm/dK}}$ yield poor
$Q$ values ($\le 0.01$) if the $N=2\,000$ are included in the fits. If we
discard the smallest lattice in these two cases, all fits shown in
Table \ref{tab:other} are characterized by $Q \ge 0.10$. The final estimate
is then
\begin{equation}
\beta/\nu = 0.51587 \pm 0.00082,
\label{eq:beta_nu}
\end{equation}
and, by using our estimate for $1/\nu$ in (\ref{eq:nu_all}),
\begin{equation}
\beta = 0.32498 \pm 0.00077.
\label{eq:beta_1}
\end{equation}
If hyperscaling is valid, the estimate (\ref{eq:beta_nu}) would imply
$\gamma/\nu = D - 2\beta/\nu = 1.9683(17)$, which however turns out to be
only barely
consistent with the direct estimate (\ref{eq:gamma_nu}) of $\gamma/\nu$.

The FSS of $d [\langle |m| \rangle]_{\rm av}/dK$ is less well behaved.
Here we had to discard for the $K_{{\rm max}_C}$, $K_{{\rm max}_{dm/dK}}$,
$K_{{\rm max}_{dlnm/dK}}$, and $K_{{\rm max}_{dlnm2/dK}}$ sequences both
the $N=2\,000$ and $N=4\,000$ data in order to guarantee that all fits
entering the average have a goodness-of-fit parameter $Q \ge 0.15$. We then
obtain
\begin{equation}
(1-\beta)/\nu = 1.0680 \pm 0.0014,
\label{eq:1-beta_nu}
\end{equation}
and by inserting the estimate (\ref{eq:nu_all}) for $1/\nu$,

\begin{equation}
\beta/\nu = 0.5194 \pm 0.0026,
\label{eq:beta_nu2}
\end{equation}
and
\begin{equation}
\beta = 0.3272 \pm 0.0014.
\label{eq:beta_2}
\end{equation}
Recent MC estimates for regular SC lattices are
$\beta/\nu = 0.518(7)$ in Ref.~\cite{fl3dis} and
$\beta/\nu = 0.5185(15)$ in Ref.~\cite{bloete1}.

\paragraph{E. Critical exponent $\alpha$:}

\begin{figure}[bhp]
\vskip 6.8truecm
\includegraphics{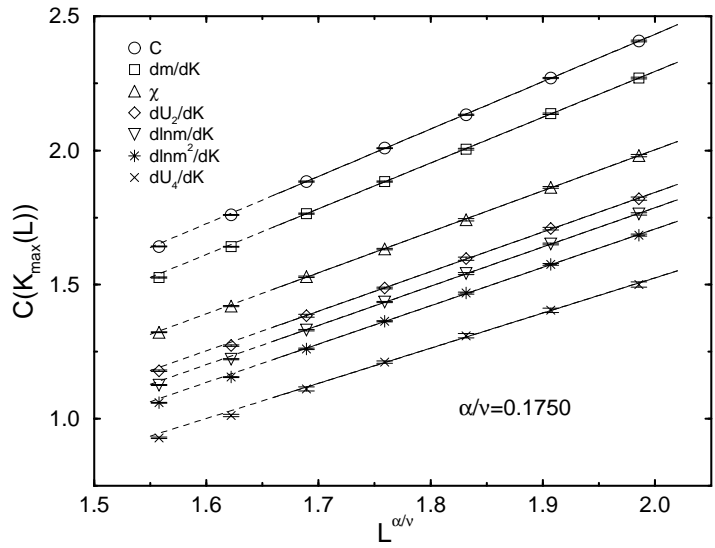}
\caption[a]{{\em FSS behavior of the specific heat, assuming
$\alpha/\nu = 2/\nu - D = 0.1750$.}}
\label{fig:C}
\end{figure}

Due to the regular background term $C_{\rm reg}$ in the FSS behavior
(\ref{eq:fss_C}), the specific heat is usually among the most difficult
quantities to analyse \cite{hj_prl}. We tried non-linear fits to the ansatz
$C = C_{\rm reg} + a L^{\alpha/\nu}$, but for most sequences of
pseudo-transition points the errors on the parameters of this fit turned
out to be large. We therefore fixed the exponent $\alpha/\nu$ at the value one
would expect if hyperscaling is valid,
%
\begin{eqnarray}
\alpha/\nu \!\!&=&\!\! 2/\nu - D = 0.1750 \pm 0.0024,\\[0.2cm]
%
\alpha \!\!&=&\!\! 2 - D \nu = 0.1102 \pm 0.0015,
\label{eq:alpha}
\end{eqnarray}
and tested if linear two-parameter fits yield acceptable
goodness-of-fit values. The results are shown in Fig.~\ref{fig:C}. We
see that over the whole range of lattice sizes the expected linear
behavior is satisfied. The quantitative analysis reveals some deviations
for the two smallest lattice sizes, but for the fits starting with
$N=8\,000$ we obtained for all seven sequences of pseudo-transition points
goodness-of-fits parameters $Q \ge 0.5$.

\paragraph{F. Binder parameters $U_2$ and $U_4$:}

\begin{figure}[bhp]
\vskip 6.8truecm
\includegraphics{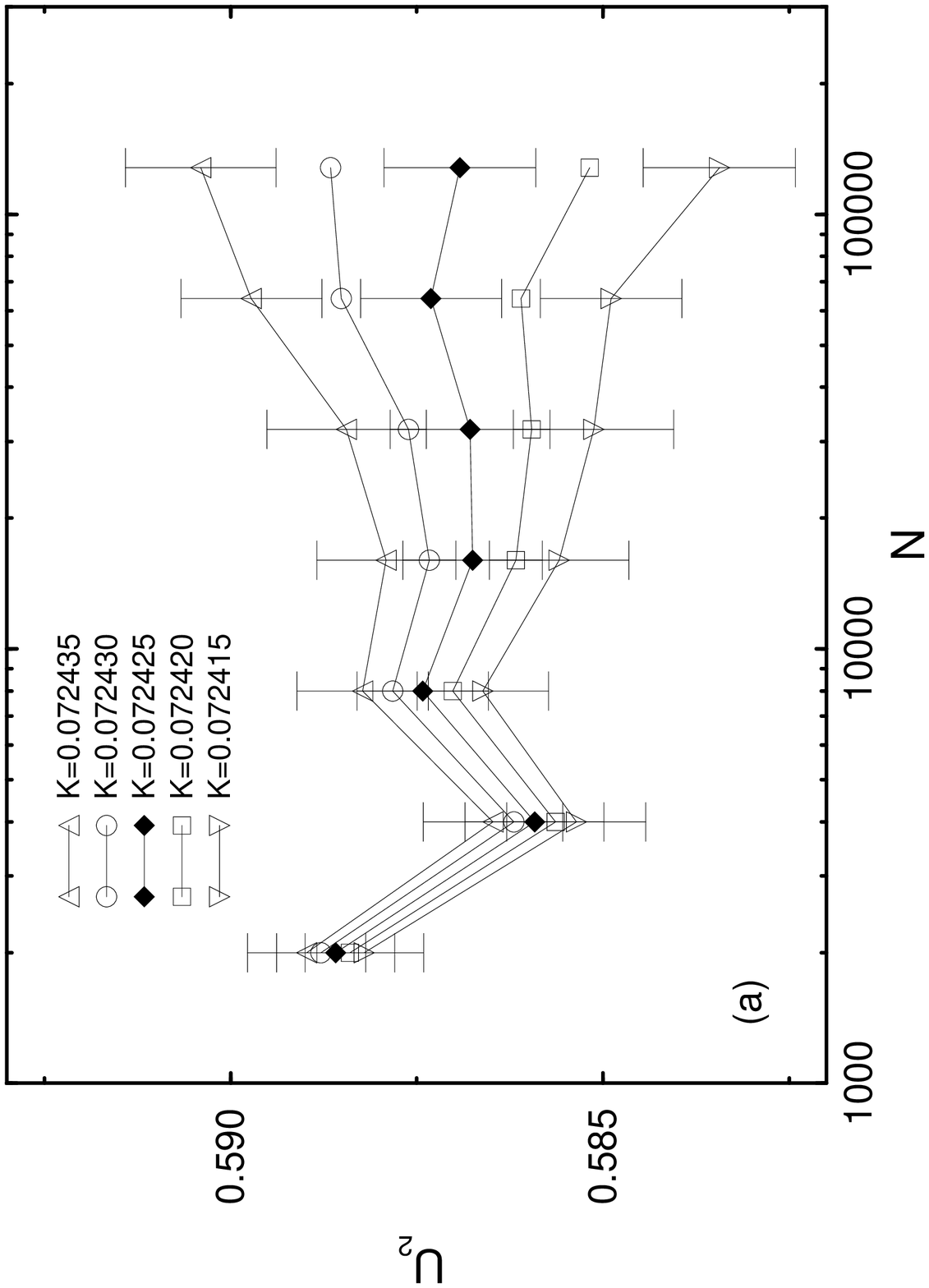}
\vskip 7.5truecm
\includegraphics{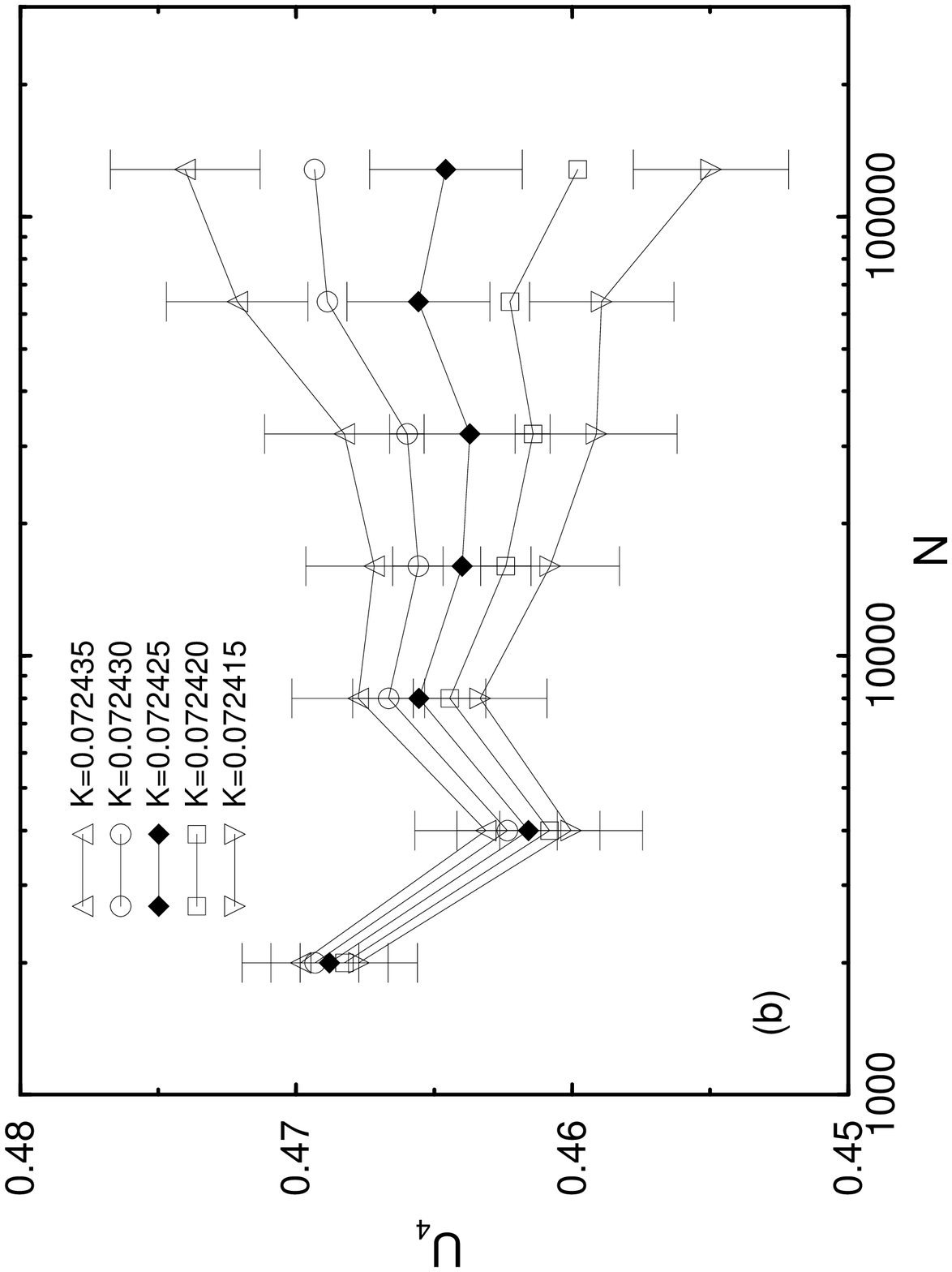}
\caption[a]{{\em FSS behavior of the magnetic cumulants. The central value
of $K$ is our best estimate (\ref{eq:kc}) for the inverse critical temperature.
For the neighboring curves the $K$ values vary by about one statistical error
bar.
}}
\label{fig:Uscal}
\end{figure}

It is well known \cite{binder81} that the $U_{2p}(K)$ curves for different
lattice sizes $L$ should intersect around ($K_c$, $U_{2p}^*$) with slopes
$U'_{2p} \equiv dU_{2p}/dK \propto L^{1/\nu}$,
where $U_{2p}^*$ is the (weakly universal) ``renormalized charge''.
In Fig.~\ref{fig:Uscal} we show $U_2$ and $U_4$ as a function of $N \equiv L^3$
for 5 $K$-values around $K_c \approx 0.072\,425$. At our best
estimate of $K_c$, both cumulants seem indeed to be almost independent
of the lattice size. Taking as final estimate the weighted mean value
(i.e., a least-squares fit to a constant) over the results for
$N=8\,000$ -- $128\,000$, we obtain
\begin{eqnarray}
U_2^* \!\!&=&\!\! 0.58706 \pm 0.00044, \\[0.2cm]
U_4^* \!\!&=&\!\! 0.4647 \pm 0.0012.
\end{eqnarray}
The variation due to the uncertainty in $K_c$ is about
twice the statistical error at fixed $K$ (0.00080 for $U_2$ and
0.0020 for $U_4$). For comparison, for the standard nearest-neighbor
Ising model on a SC lattice, Ferrenberg and Landau \cite{fl3dis} estimated
$U_4^* \approx 0.47$, by combining results for three different spin
models belonging to the Ising universality class,
Bl\"ote {\em et al.\/} \cite{bloete1} derived $U_4^* = 0.4652(4)$,
and Ballesteros {\em et al.\/} obtained $U_4^* = 0.4656(4)[4]$. For
the $\phi^4$ lattice field theory Hasenbusch \cite{hasenbusch}
extracted $U_4^* \approx 0.46555(9)$.
%
%
     \section{Concluding remarks}    \label{sec:conclus}
%
We have performed a detailed finite-size scaling analysis of single-cluster
Monte Carlo simulations of the Ising model on three-dimensional Poissonian
random lattices of Voronoi/Delaunay type.
At first sight our use of different quantities to estimate the same critical
exponent might appear redundant, since the
various estimates are, of course, not independent in a statistical sense. Their
consistency, however, gives confidence that corrections to the asymptotic
scaling behavior are very small and can safely be neglected.
Our estimates for the exponents $\nu$, $\beta/\nu$, and $\gamma/\nu$
are all consistent with the best numerical estimates for the
three-dimensional Ising model and $\phi^4$ field theory on regular
lattices -- at a very high level
of accuracy which is comparable with the best estimates coming from
field theoretical techniques, cf.\ Table~\ref{tab:3dexpos} \cite{footnote2}.
While our exponent ratio $\gamma/\nu$ would also be compatible with recent
estimates for the 3D Ising disordered fixed point, our estimate for
$U_4^*$ is more consistent with the pure Ising model estimates. The
cleanest result yields the critical exponent $\nu$, where
our result agrees within error bars with all previously
derived estimates for the pure model but is clearly incompatible with the
disordered fixed point value.
\begin{table}[hb]
\caption[a]{{\em Recent estimates of critical parameters of the pure and
disorderd 3D Ising model
(SC = simple cubic lattice, RG = renormalization group, SD = site-dilution,
RIM = random Ising model).}}
\label{tab:3dexpos}
\vspace{3ex}
\begin{center}
\begin{tabular}{|l|lll|}
\hline
\hline
 \multicolumn{1}{|c|}{method}      &
 \multicolumn{1}{c}{$\nu$}          &
 \multicolumn{1}{c}{$\gamma/\nu$}   &
 \multicolumn{1}{c|}{$U_4^*$}        \\
\hline
SC \cite{fl3dis}              & 0.6289(8)    & 1.970(14)
    & 0.47       \\
SC \cite{bloete1}             & 0.6301(8)    & 1.9630(30)
    & 0.4652(4)  \\
SC \cite{Ba_etal99}         & 0.6294(5)[5] & 1.9626(6)[6]
    & 0.4656(4)[4] \\
SC $\phi^4$ \cite{hasenbusch} & 0.6296(3)[4] & 1.9642(4)[5]
    & 0.46555(9) \\
RG \cite{Guida1998}           & 0.6304(13)   & 1.966(6) & $-$ \\
\hline
This work                          & 0.6299(5)    & 1.9576(13)
    & 0.4647(12) \\
\hline
SD SC \cite{Ba_etal98}        & 0.6837(24)[29] & 1.9626(36)[9]
    & 0.449(5)[2] \\
RIM-RG \cite{Pelissetto2000}  & 0.678(10)      & 1.970(3)
    & $-$ \\
RIM-RG \cite{holo00}          & 0.675          & 1.951
    & $-$ \\
\hline
\hline
\end{tabular}
\end{center}
\end{table}
We thus obtain strong evidence that,
for the considered lattice sizes up to
$N=128\,000 \approx 50^3$, the Ising model on three-dimensional Poissonian
random lattices of Voronoi/Delaunay type behaves effectively as on regular
lattices.

Of course, we cannot exclude the possibility
that on much larger length scales (lattice sizes) the scaling behavior
may change. Such a late crossover is conceivable in the case of weak
disorder, where the asymptotic critical behavior governed by a ``disordered''
fixed point may show up only in the extremely close vicinity of criticality,
that is at extremely large system sizes in a finite-size scaling analysis.
Even though the qualitative scaling behavior is expected to be universal,
quantitative properties of the crossover point such as its location should
depend on the strength of the disorder via non-universal amplitudes.
In order to obtain for the random lattices a rough estimate of the strength
$\cal S$ of the local connectivity disorder we have computed
the relative variance of the local coordination numbers which may be viewed as
a measure for the size of effective temperature variations over the lattice.
From the probability density $P(q)$ displayed in Fig.~\ref{fig:P(q)} it is
straightforward to obtain
\begin{equation}
{\cal S} \equiv \overline{(q - \overline{q})^2}/\overline{q}^2 = 0.0461,
\label{eq:calS_q}
\end{equation}
with $\overline{q} = 2 +48\pi^2/35 = 15.5354\dots$. Similarly, for
two-dimensional Poissonian Voronoi/Delaunay random lattices one finds
${\cal S} = 0.0491$ with $\overline{q} = 6$. The relative variance
(\ref{eq:calS_q}) can be compared with the fluctuations of the number $B$ of
active bonds per site in bond-diluted models. Here $B$ follows a
binomial distribution,
$P(B) = \left(\!\!\scriptsize \begin{array}{cc} 2D\\B\end{array}\!\!\right)
p^B (1-p)^{2D-B}$, where $D$ is the dimension and $p$ denotes the probability
for a bond to be active (such that $p=1$ corresponds to the pure model), and
one obtains
\begin{equation}
{\cal S} = \overline{(B - \overline{B})^2}/\overline{B}^2 = \frac{1}{2D}
\frac{1-p}{p},
\label{eq:calS_B}
\end{equation}
with $\overline{B} = 2D p$. By equating Eqs.~(\ref{eq:calS_q}) and
(\ref{eq:calS_B}) and solving for the dilution parameter $p$ one
can thus determine an associated bond-dilution model with the same
local disorder fluctuations as for the random lattices. For the
three-dimensional case this yields $p = 0.7834$, and in two
dimensions one finds $p=0.8358$. In the terminology of
three-dimensional bond-diluted Ising \cite{BCBJ_01,BCBJ_02,HJ_01}
and $q$-state Potts \cite{CBJB_01,HJ_lat01} models such a value of
$p$ belongs to the weak dilution regime where some influence of
the disordered fixed point can be observed, but it is still
difficult to clearly disentangle it. For site-diluted models the
corresponding $p$-value is presumably higher, in particular for
weak dilution, since all bonds around a vacant site are
non-active. In the latter models, of course, the dilution
parameter $p$ can easily be tuned to study more accessible
regions.

In view of the very high quality of our fits based on the leading
FSS ansatz only we must conclude that in the case of
Voronoi/Delaunay random lattices very much larger system sizes
would be necessary to observe the expected crossover to the
critical behavior associated with the disordered fixed point. This
was clearly outside the scope of the present study and its
computer budget which was equivalent to several years of fast
workstation CPU time. Instead of further increasing the system
size, an alternative and more promising route for future studies
could be a systematic variation of the random lattice construction
by modifying the Poissonian nature of the site distribution such
as to achieve larger values of $\cal S$ (corresponding to smaller
values of $p$), or an investigation of the present random lattices
coupled to a model with a larger critical exponent $\alpha$ where
the expected crossover should set in for reasonable lattice sizes
already for a moderate degree of local disorder.
%
%
\section*{Acknowledgements}
This work was partially supported by the NATO collaborative
research grant CRG 940135. The numerical simulations were
performed on a T3D parallel computer of Zuse-Zentrum f\"ur
Informationswissenschaften Berlin (ZIB) under grant No.\ bvpf07.
WJ would like to thank Kurt Binder for useful discussions and
acknowledges support from the Deutsche Forschungsgemeinschaft
(DFG) through a Heisenberg fellowship in an early stage of the
project. He also wishes to thank Amnon Aharony, Eytan Domany and
Shai Wiseman for helpful discussions on disordered systems
during visits supported by the German-Israel-Foundation (GIF)
under contracts No.\ I-0438-145.07/95 and No.\ I-653-181.14/1999.
He is also partially supported by the EC IHP Network grant
HPRN-CT-1999-00161: ``EUROGRID''. RV acknowledges partial support
by CICYT under contract AEN95-0882. \clearpage
\newpage

\appendix

\section*{Appendix: Random lattice construction}
The employed algorithm for the random lattice construction works as
follows. Adapting the
method described in Ref.~\cite{Tanemura}, we first draw randomly $N$
sites uniformly distributed in a unit volume, thereby approximating a
Poissonian distribution. For alternative distributions discussed in the
literature see, e.g., Refs.~\cite{dla,caer}. In the second step we link the
sites according to the Voronoi/Delaunay prescription. We start by picking
the first site that we
drew and locate all its nearest neighbors, keeping them stored in an array.
Then we proceed to the second site and search for all its nearest neighbors
too; once finished with the second site we keep repeating the procedure until
we have done it for all the sites. Of course, with this method we locate a
given link twice, but the simplicity of its implementation pays off.

Starting from a given site, $x_1$, the linking procedure works as follows.
Its nearest neighbor, $x_2$, is
located from within the few hundred sites forming, or belonging to,
the ``cloud-of-neighbors"
around $x_1$. We will comment later the issue of how to determine a
cloud-of-neighbors for a given site. Notice that some care must be exercised
when approaching the boundaries of the
lattice to ensure the periodic boundary conditions. Afterwards a third site,
$x_3$, is searched for, and a triangle is constructed. In order to locate this
third site, we draw circumferences going through $x_1$, $x_2$, and the
few hundred sites belonging to the cloud-of-neighbors of $x_1$.
We pick as $x_3$ the site for which the radius of the
circumference is the smallest. From the triangle $\bigtriangleup(x_1,x_2,x_3)$
we proceed to locate a fourth site, $x_4$, linked to $x_1$ and build a
tetrahedron, $\tau(x_1,x_2,x_3,x_4)$. When a triangle still has not been used
to build a tetrahedron from it, it is termed ``active" and a logical flag is
turned ``on''. When already used, it is renamed to ``inactive" and the flag
is turned ``off''.

To construct a tetrahedron from a triangle we split the volume in two half
spaces: one ``above'' and the other ``below'' the plane lying on the
triangle $\bigtriangleup(x_1,x_2,x_3)$. Let us suppose that we search for
$x_4$ in the half space ``above'' the triangle. In order to determine $x_4$,
we draw spheres going through $x_1$, $x_2$, $x_3$ and the sites
belonging to the cloud-of-neighbors of $x_1$ placed in the half space
in which we are working. If we happen to find
several trial sites for which their distance to the circumcenter of the
triangle $\bigtriangleup(x_1,x_2,x_3)$ is smaller than the radius of the
circumscribed circle of $\bigtriangleup(x_1,x_2,x_3)$, then we pick as $x_4$
the site for which the radius of the circumscribed sphere of
$\tau(x_1,x_2,x_3,x_4)$ is the biggest. If, on the contrary, all the trial
sites lie at a distance from the circumcenter of the triangle
$\bigtriangleup(x_1,x_2,x_3)$ greater than the radius of the circumscribed
circle of $\bigtriangleup(x_1,x_2,x_3)$, then we pick as $x_4$ the site for
which the radius of the circumscribed sphere of $\tau(x_1,x_2,x_3,x_4)$ is the
smallest. From the newly built tetrahedron $\tau(x_1,x_2,x_3,x_4)$, we can
take two ``active" triangles, $\bigtriangleup(x_1,x_2,x_4)$ and
$\bigtriangleup(x_1,x_3,x_4)$ to continue building tetrahedra from triangles,
and then triangles from tetrahedra. The closer neighbors of a given site
$x_1$ are all found when there is no ``active'' triangle left connected to
the site.

When describing how to locate $x_1$'s nearest neighbor, $x_2$, or
how to find $x_3$ afterwards we emphasized that we only search from within
the sites forming a cloud-of-neighbors around $x_1$. Its meaning is that
{\em before\/} starting the linking procedure we set up an array for each
site in the lattice containing the sites forming its cloud.
A given site will belong to the cloud of, say, $x_1$ if it lies within a
sphere centered in $x_1$. The radius of the sphere is chosen such that,
on the average, the
number of sites within the sphere is three times of an {\em a priori\/} upper
limit to the maximum number of links that a site is likely to have in a
finite Voronoi/Delaunay random lattice.
To implement an efficient search of the sites which will belong to
the cloud-of-neighbors of a given site, we subdivided the unit volume into
smaller boxes. The optimal box size should be large enough to ensure
that nearest neighbors will be located in the same box or at least in one of
the 26 surrounding boxes, but small enough to minimize the time needed for
testing all trial sites in a box.
%
     
%
\newpage
%
%

%
%
%
%
%
%
%
\clearpage\newpage
\clearpage \newpage
\clearpage \newpage
\clearpage \newpage


\begin{thebibliography}{99}
\bibitem{foot1_sytematic}
In Refs.~\cite{Ba_etal99} and \cite{hasenbusch} the error bars are given
separately for purely statistical and more systematic errors.
\bibitem{footnote2}
In the summarizing Table~\ref{tab:3dexpos} we confine ourselves to those
critical parameters which are usually measured or calculated independently.
Other exponents are often derived by scaling or hyperscaling relations which
does not add any new information.
\end{thebibliography}
\end{document}